\documentclass[12pt]{article}
\newtheorem{theorem}{Theorem}
\newtheorem{lemma}{Lemma}
\newtheorem{proposition}{Proposition}
\newtheorem{definition}{Definition}
\newtheorem{example}{Example}
\newcommand{\blackslug}{\mbox{\hskip 1pt \vrule width 4pt height 8pt 
depth 1.5pt \hskip 1pt}}
\newcommand{\QED}{\quad\blackslug\lower 8.5pt\null\par\noindent}
\newcommand{\proof}{\par\penalty-100\vskip .5 pt\noindent{\bf Proof\/: }}

\title{Truth Revelation in Approximately Efficient Combinatorial Auctions}
\author{Daniel Lehmann \\ School of Computer Science and Engineering, \\
Hebrew University, Jerusalem 91904, Israel \and Liadan Ita O'Callaghan \and Yoav Shoham \\ Robotics Lab., Computer Science Dept., Stanford University}
\date{Keywords: Combinatorial Auctions, Mechanism Design, Computational Complexity}
\begin{document}
\maketitle

\begin{abstract}
Some important classical mechanisms 
considered in Microeconomics and Game Theory
require the solution of a difficult optimization problem. 
This is true of mechanisms for combinatorial auctions, 
which have in recent years assumed practical importance, 
and in particular of the gold standard for combinatorial auctions, 
the Generalized Vickrey Auction (GVA).
Traditional analysis of these mechanisms - in particular, their truth
revelation properties - assumes that the optimization problems are
solved precisely. 
In reality, these optimization problems can usually be
solved only in an approximate fashion. 
We investigate the impact on such mechanisms 
of replacing exact solutions by approximate ones.
Specifically, we look at a particular greedy optimization method. 
We show that the GVA payment scheme does not provide 
for a truth revealing mechanism. 
We introduce another scheme that does guarantee truthfulness 
for a restricted class of players. 
We demonstrate the latter property by identifying natural
properties for combinatorial auctions and showing that, for our restricted class
of players, they imply that truthful strategies are dominant.
Those properties have applicability beyond the specific auction studied.
\end{abstract}



\section{Introduction}
\label{sec:intro}
This articles concerns combinatorial auctions (also called
combinational), that is, auctions in which multiple goods are available
and in which bidders can post bids for subsets, i.e.\ bundles, of the goods. 
Such auctions
have become the object of increased interest recently, in part because
of the general interest in auctions, and in part because of specific
auctions in which combinatorial bidding would seem natural, 
such as the series of the FCC spectrum 
auctions~\cite{McMillan:Spectrum,Cramton:97,Milgrom:98}.
\footnote{Up until now the FCC auctions have not in fact been
combinatorial, due in part to the complexity problem discussed below.
However,
the FCC is currently actively considering a combinatorial
auction.}

Combinatorial auctions (henceforth CAs) typically require the solution
of one or more difficult optimization problems. The computational
complexity of these problems threatens to render the traditional auction
designs a mere theoretical construct. One approach to meeting this
threat is to replace the exact optimization by an approximate one. 
This, however, gives rise to a new challenge: 
traditional analysis of established CA mechanisms relies strongly on the fact
that the goods are allocated in an optimal manner, and
the properties guaranteed by the mechanism (such as truthful bidding, to
be defined later), disappear if the allocation is anything
less than optimal. 
This is true in particular of the Generalized Vickrey Auction (GVA), 
also defined later, which is widely taken to be the gold standard for CAs. 
The primary focus of this article
is to present a simple approximate optimization method for CAs that
possesses two attractive properties: 
\begin{itemize}
\item the method performs a reasonably effective optimization, and 
\item there exists a novel payment scheme
which, when coupled with the approximate optimization method, makes for
a combinatorial auction in which truth-telling is a dominant strategy.
\end{itemize}
In order to show the latter property we identify several axioms which
are sufficient to ensure truth-telling for a restricted class of players,
in any combinatorial auction;
these axioms are interesting in there own right, as they can be applied
to auctions other than the one discussed here.

{\em Note}: Since we aim to make this article easily accessible to 
both computer scientists and game theorists, 
we include some rather basic material.

\section{A brief introduction to combinatorial auctions}
\label{sec:combauctions}
In this section we briefly cover the notions of complementarity 
and substitutability, as motivating CAs; 
the two degrees of freedom in a sealed-bid CA, namely
allocation and payment policies; 
and why one needs to be careful when
applying the desiderata of efficiency and revenue maximization to CAs.


\subsection{Complementarity and substitutability}
\label{sec:compl}
Throughout this article we shall consider single-side CAs with a single
seller and multiple buyers. 
The reverse situation with a single buyer and multiple sellers is symmetric; 
the two-sided case, with multiple buyers and sellers, is more complex, 
and lies outside the scope of this article. 
Let us assume, then, that an auctioneer is selling a number of different
goods.
In such a situation, a bidder may be willing to pay more for the whole
than the sum of what he is willing to pay for the parts: 
this is the case if the parts complement each other well, 
e.g., a left shoe and a right shoe.
This phenomenon is called {\em complementarity}.
In other cases, a bidder may be willing to pay
for the whole only less than the sum of what he is willing
to pay for the parts, maybe only as much as one of the parts.
This is especially the case if the bidder has a limited budget
or if the goods are similar, or interchangeable, e.g., two tickets
to the same performance.
This phenomenon is called {\em substitutability}.
In general, complementarity and substitutability can both play heavily
in the same auction.

In the absence of complementarity and substitutability,
i.e.\ if every participant values a set of goods at the sum of
the values of its elements, one should organize the multiple
auction as a set of independent simple auctions,
but, in the presence of complementarity,
organizing the multiple auction as a set
or even a sequence
of simple auctions will lead to less than optimal results:
e.g, a participant ending up with a left shoe and another one with
the right shoe, or the left shoe auctioned for almost nothing because
bidders fear not to be able to get the right shoe and the right shoe
then
auctioned for nothing to the buyer of the left shoe since no one is
interested in just a right shoe. The problem is particularly acute when
the complementarity and substitutability relations vary among the
various bidders.

\subsection{Specifying a combinatorial auction}
\label{sec:specifying}
Several auction designs have been proposed to deal with complementarity
and substitutability. The Simultaneous Ascending
Auction was devised in connection with the FCC Spectrum Auction 
mentioned above,
but its discussion is beyond the scope of this paper. 
In this paper, we shall consider only
what is perhaps the most obvious approach, which is to allow
combinatorial bidding.
For the history of combinatorial auctions, see~\cite{Rothkopf:83}.
What does it take to specify a CA? 
In general, any auction must specify three elements: 
the bidding rules (that is, what one is allowed to bid for and when), 
the market clearing rules (that is, 
when is it decided who bought what and who pays what), 
and the information disclosure rules (that is, 
what information about the bid state is disclosed to whom and when).

We will be considering only one-stage, sealed-bid CAs; in these, each
bidder submits zero or more bids, the auction clears, and the results
are announced. The third element of the specification is thus
straightforward: no information is released about other bidders' bids
prior to the close of the auction.

The first element of the specification is almost as straightforward:
each bidder may submit one or more bids, each of which mentions a subset
of the goods and a price. 
One has to be precise, however, about the semantics
of the collection of bids submitted by a single bidder; 
if I bid \$5 for $a$ and \$7 for $b$, what does it mean about my willingness 
to pay for $\{a,b\}$?
If I bid \$10 for $\{a,b\}$ and \$20 for $\{b,c\}$, 
what does it mean about my willingness to pay for $\{a,b,c\}$? 
This is not a mysterious issue, but one needs to be precise about it. 
We shall return to this issue later 
when we discuss the notion of a bidder's {\em type}.

The scheme above allows one to express complementarity.
Bidding for \$5 for $a$, \$7 for $b$ and \$15 for $\{a,b\}$ clearly
indicates complementarity.
On the face of it, though, substitutability cannot be expressed,
since bidding \$8 for $\{a,b\}$, \$5 for $a$ and \$7 for $b$ does
not preclude, under the usual market clearing rules, one being allocated
$a$ and $b$ separately. 
However, a simple encoding trick presented in~\cite{FLBS:99} 
allows the expression
of substitutability, at least partially.

Thus, so far the designer of a combinatorial auction has no discretion. 
Only the second element of specification, the clearing policy, 
provides choices.
There are two choices to be made here: which goods does every bidder
receive, and how much does every bidder pay? 
We address these below.

\subsection{Maximizing efficiency and revenue}
\label{sec:maximizing}
The standard yardsticks for auction design, which are sometimes at odds
with one another, are guaranteeing efficiency and maximizing (in our case,
the seller's) revenue. 
We shall be concentrating primarily on efficiency in this article,
but a very preliminary study of revenue is found in 
section~\ref{sec:evaluation}. Efficiency means that the allocation
(of goods and money) resulting from the auction is Pareto optimal:
no further trade among the buyers can improve the situation of some 
trader without hurting any of them.
This is typically achieved by ensuring that the clearing rules maximize
the sum of the values the
various bidders place on the actual allocation decided on by the
auctioneer. 
On the whole, one can expect that an efficient auction, after which
the participants are globally satisfied, allows the seller to extract
a higher revenue than an inefficient auction after which the level
of social satisfaction is lesser. 
Efficiency, therefore, which may be a goal
in itself, may also be a step in the direction of revenue maximization.
In fact, this correlation holds only in part and auctions that are maximizing
revenue are not always efficient~\cite{Myer:Opt}.
Nevertheless, we shall seek efficient, at least approximately, 
auction mechanisms.

Note four problems here. 
We have already
mentioned that bidders specify bids, not their profile of preferences
over bundles. 
This does not pose a real challenge, 
so long as one is clear about the meaning of those bids.
The second one is that those profiles of preferences over bundles
do not allow for a full specification of preferences about the outcomes
of the auction, i.e.\ the resulting allocation. 
A bidder cannot express {\em externalities}, e.g. that he would prefer,
if he does not get a specific good, this good to be allocated to bidder $X$ and
not to bidder $Y$. 
Third, we have an optimization problem on our hand; as it turns out, it
is an NP-hard optimization problem that cannot be even approximated
in a feasible way, in the worst case.
This means that for all practical purposes there does not exist
a polynomial-time algorithm for computing the optimal allocation, or
even for computing an allocation that is guaranteed to be off from
optimal by at most a constant, any given constant.
The fourth and deepest problem is that the optimization is supposed to
happen over the bidder's true valuations, as opposed to merely their bid
amounts, but that information is not available to the auctioneer
and the bidder will reveal this information only if it is in his/her 
best interest.

An ingenious method, discussed in the next section, has been developed
in game theory to overcome the fourth problem. 
The problem is that not only does it not address the second problem, 
it actually mildly exacerbates it
by requiring that the optimization be performed once for each bidder.
The primary goal of this paper is to devise a method which promises
good (albeit sub-optimal) efficiency, while being computationally
feasible. 
In a nutshell, the goal is to simultaneously ensure economic and
computational efficiency.

\section{Mechanism design for CA}
\label{sec:mechanism}
In this section, we consider the design of combinatorial
auctions as a problem of designing a game of incomplete information
for which the weakly-dominant strategies present a {\em good} way
of allocating the goods and paying for them.
The general setting is that of 
economic mechanism design: see~\cite[Chapter 23]{MasAndal:Micro}, for example, 
for an introduction to the field and~\cite{Varian:95} for a description of
auctions in this framework.
Contrary to the latter, we shall restrict our description to
combinatorial auctions in which no externalities can be expressed.
Informally, each bidder sends a message describing (truthfully
or not) his preferences, the auctioneer, then, computes the resulting
allocation of the goods and the payments, 
based on the bidders' messages but according
to rules known in advance. The mechanism is a {\em truthful} one
if it is in the best interest of the bidders to send messages that 
truthfully reveal their preferences.

Formally, we consider a set $P$ of $n$ bidders. 
The indices $i$, $j$ \mbox{$1 \leq i , j \leq n$},
will range over the bidders.
Bidders are selfish, but rational, and trying to maximize their
utility in the final outcome.
A bidder knows his own utility function, i.e.\ his {\em type},
but this information is private and neither the auctioneer nor the other
players have access to it. 
The final result of an auction consists of two elements:
an allocation of the goods and a vector of payments from the bidders to
the auctioneer, both of which are functions of the bidders' declarations,
i.e.\ bids. 
Formally, we have a finite set $G$ of $k$ goods and an allocation
is a {\em partial} function from $G$ to $P$, i.e.\
a function \mbox{$a : G \rightarrow P'$}, with 
\mbox{$P' = P \cup \{unallocated\}$},
since we do not insist that all goods be allocated.
Notice that the allocations produced by the Generalized Vickrey Auctions
of section~\ref{sec:GVA} and by our Greedy algorithm~\ref{sec:greedy}
are not always total.
The set of outcomes, i.e.\ allocations, is ${\cal O} = P'^{G}$,
the set of partial functions from $G$ to $P$.
Since we do not allow for externalities, 
the set $\Theta_{i}$ of the possible types for bidder $i$
is ${{\bf R}_{+}}^{2^{G}}$, where ${\bf R}_{+}$ is the set of all non-negative
real numbers. 
Notice that such a type allows for both complementarity and substitutability,
but not for externalities.
Since the set $\Theta_{i}$ does not depend on $i$, we shall write $\Theta$.
An element of $\Theta$ assigns a real non-negative valuation to every
possible bundle.
The set $\Theta$ is also the set of messages that bidder
$i$ may send. 
A bidder may send any element of $\Theta$, irrespective
of his (true) type, i.e.\ a bidder may lie.
We shall typically use $t$ to denote a (true) type, $d$ to denote a message,
$T$ or $D$ to denote
vectors of $n$ types and $P$ for a payment vector, i.e.\ a vector of $n$
non-negative numbers.

Since we assume the Independent Value Model and Quasi-Linear utilities,
fairly standard assumptions in the field,
the utility, for a bidder of type $t$, of bundle $s$ and payment $x$
is:
\begin{equation}
\label{eq:utility}
u = t(s) - x
\end{equation}

\begin{definition}
\label{def:mech}
A (direct) mechanism for combinatorial auctions consists of
\begin{itemize}
\item an allocation algorithm $f$ that picks, for each vector $D$
($D$ is a vector of declared types), an
allocation $f(D)$,
\item a payment scheme $p$ that determines, for each vector $D$ 
a payment vector \mbox{$p(D)$}: $p_{i}(D)$ is paid by bidder $i$ to
the auctioneer. 
\end{itemize}
\end{definition}
Let us denote the bundle obtained by $i$ as:
\begin{equation}
\label{eq:gi}
g_{i}(D) = {f(D)}^{-1}(i)
\end{equation}
{\em Notation}: In general $g_{i}$ depends on 
the allocation algorithm $f$, but when $f$ is clear from the context
we shall abuse the notation and
treat $g_i$ as a direct function of the bid vector, $D$.
Equation~\ref{eq:utility} implies that if bidder $i$ has (true) type $t$,
his utility from the mechanism is: 
\begin{equation}
\label{eq:utility2}
u_{i} = t(g_{i}(D)) - p_{i}(D), 
\end{equation}
where
\mbox{$D = \langle d_{1} , \ldots , d_{n} \rangle$}
is the vector of declarations. 

The first term of this sum, $t(g_{i}(D))$ is often called the valuation
of $i$: $v_{i}(f(D) , t)$.
The game-theoretic solution concept used throughout this paper is
that of a weakly-dominant strategy,
that is a strategy that is as good as
any other for a given player, no matter what other players do.
This is in contrast with the weaker and more common
notion of Nash equilibria. The particular property we would
like to ensure for our mechanism is that the dominant strategy
for each player is to bid his true valuation; in other words, no
bidder can be better-off by lying, no matter how other bidders behave. 
This is obviously a very strong requirement.

A mechanism is truthful if no bidder can be better-off by lying,
even if other bidders lie. This is a very strong requirement, 
making for a very sturdy mechanism.
\begin{definition}
\label{def:truthful}
A mechanism \mbox{$\langle f , p \rangle$} is truthful if and only if for every 
\mbox{$i \in P$}, \mbox{$t \in \Theta$} and any vector $D$ of declarations,
if $D'$ is the vector obtained from $D$ by replacing the $i$-th coordinate
$d_{i}$ by $t$, then:
\mbox{$t(g_{i}(D')) - p_{i}(D') \geq t(g_{i}(D)) - p_{i}(D)$}.
\end{definition}
In the definition above, $t$ is the true type of bidder $i$ and $D$ is
a vector of declared types.
The term $t(g_{i}(D))$ represents the true satisfaction $i$ receives from
the allocation resulting from declarations $D$
and $t(g_{i}(D'))$ represents his true satisfaction from the allocation
that would have been obtained had $i$ been truthful.
\section{The generalized Vickrey auction}
\label{sec:GVA}


A very general method for design truthful mechanisms has been
devised by Clarke and Groves~\cite{Clarke:71,Groves:73}.
Applied to combinatorial auctions it generalizes the
second price auctions of Vickrey~\cite{Vickrey:61}.
We shall now describe those generalized Vickrey auctions, 
prove that the mechanism described is truthful and then discuss the
complexity issues that render those auctions unfeasible when $k$,
the number of goods, is large.
Generalized Vickrey Auctions (GVAs) appear to be part of the
folklore of mechanism design. A description of a more general type
may be found in~\cite{VarianMacK:GVA,Varian:95};
we adopt a special case of it, one which does not allow for externalities.

In a GVA, the allocation chosen maximizes
the sum of the declared valuations of the bidders, 
each bidder receives a monetary amount that equals the sum of the declared 
valuations of all other bidders, 
and pays the auctioneer the sum of such valuations that
would have been obtained if he had not participated in the auction.
A way to describe such an auction, in which $i$ does not participate, 
is to consider the auction
in which bidder $i$ declares a zero valuation for all possible bundles.
A bidder with zero valuation for all bundles has no influence on the
outcome.

Formally, given a vector $D$ of declarations, the generalized Vickrey auction
defines the allocation and payment policies as follows
(notice that \mbox{$a^{-1}(i)$} is the bundle allocated to $i$
by allocation $a$, and that \mbox{$g_{i}$} is defined in 
Equation~\ref{eq:gi}):
\begin{equation}
\label{eq:f}
f(D) = {\rm argmax}_{a \in {\cal O}} \sum_{i = 1}^{n} d_{i}(a^{-1}(i)),
\end{equation}
\begin{equation}
\label{eq:p}
p_{j}(D) = - \sum_{i = 1 , i \neq j}^{n} d_{i}(g_{i}(D)) + 
\sum_{i = 1 , i \neq j}^{n} d_{i}(g_{i}(Z))
\end{equation}
where \mbox{$Z_{i} = D_{i}$} for any \mbox{$i \neq j$}
and \mbox{$Z_{j}(s) = 0$} for any bundle \mbox{$s \subseteq G$}.
Since \mbox{$d_{j}(g_{j}(Z)) = 0$}, we may as well have written:
\begin{equation}
\label{eq:p2}
p_{j}(D) = - \sum_{i = 1 , i \neq j}^{n} d_{i}(g_{i}(D)) + 
\sum_{i = 1}^{n} d_{i}(g_{i}(Z))
\end{equation}
A proof of the truthfulness of the Clarke-Groves-Vickrey mechanism may be 
found, for example in~\cite[Proposition 23.C.4]{MasAndal:Micro}.
We include the proof here only to stress how easy it is.
\begin{theorem}
\label{the:CGV}
The generalized Vickrey auction is a truthful mechanism.
\end{theorem}
\proof
Assume \mbox{$j \in P$}, \mbox{$t \in \Theta$}, $D$ is a vector
of declarations, and \mbox{$D'_{i} = D_{i}$} for any $i \neq j$
and \mbox{$D'_{j} = t$}.
By Equation~\ref{eq:f}, 
\[
\sum_{i = 1}^{n} d'_{i}(g_{i}(D')) \geq \sum_{i = 1}^{n} d'_{i}(g_{i}(D)).
\]
But, for \mbox{$E = D , D'$} we have:
\[
d'_{i}(g_{i}(E)) = d_{i}(g_{i}(E)) , {\rm \ if \ } i \neq j \ {\rm and} \ 
d'_{j}(g_{j}(E)) = t(g_{j}(E)).
\]
Therefore,
\[
t(g_{j}(D')) - p_{j}(D') + \sum_{i = 1 , i \neq j}^{n} d_{i}(g_{i}(Z))
\geq 
t(g_{j}(D)) - p_{j}(D) + \sum_{i = 1 , i \neq j}^{n} d_{i}(g_{i}(Z))
\]
and \mbox{$t(g_{j}(D')) - p_{j}(D') \geq t(g_{j}(D)) - p_{j}(D)$}.
\QED
Notice that the second term in the payment of $j$ does not depend
on $j$'s declaration and is therefore irrelevant to his decision on
what to declare.
A feature of the GVA is that no truthful bidder's utility 
can be negative.
\begin{proposition}
\label{prop:GVA}
If $j$ is truthful, his utility $u_{j}$ in the GVA is non-negative.
\end{proposition}
\proof
By Equation~\ref{eq:utility2}, since $j$ is truthful, 
by Equation~\ref{eq:p2} and finally by Equation~\ref{eq:f}:
\[
u_{j} = d_{j}(g_{j}(D)) + \sum_{i = 1 , i \neq j}^{n} d_{i}(g_{i}(D)) -
\sum_{i = 1}^{n} d_{i}(g_{i}(Z)) =
\sum_{i = 1}^{n} d_{i}(g_{i}(D)) -
\sum_{i = 1}^{n} d_{i}(g_{i}(Z)) \geq 0
\]
\QED
Since bidders truthfully declare their type and the allocation maximizes
the sum of the declared utilities, in a GVA,
the allocation maximizes the sum of the true valuations of the bidders,
i.e.\ the social welfare.
In a quasi-linear setting, this is equivalent to Pareto optimality. 
Therefore a GVA is Pareto optimal.
The mechanism to be presented in section~\ref{sec:positive}
only approximately maximizes the sum of the true valuations
of the bidders, and is not Pareto optimal.
 
As we discuss in the following sections, it is known that
algorithmic complexity considerations
imply that Pareto optimality cannot
be feasibly attained.
Specifically, ensuring Pareto efficiency requires solving
an intractable optimization problem. 
This is true even if we restrict the class of bidders severely,
as we propose in section~\ref{sec:single}.

\section{Single-minded bidders}
\label{sec:single}
As is customary, we shall consider that any algorithm 
whose running-time is polynomial in $k$ and $n$ is feasible, but any algorithm
whose running-time is not polynomial in $k$ or in $n$ is unfeasible.
The size of the set ${\cal O}$ of allocations is exponential in $k$,
if there are at least two bidders,
and the set $\Theta$ of possible types is doubly-exponential in $k$.
Since, in a direct mechanism (we consider no others), 
the message that a bidder sends describes one specific element (type)
of $\Theta$, a bidder
needs an exponential number of bits to describe his type:
the length of the messages sent in any such mechanism, 
and in a generalized Vickrey auction, is exponential in $k$, 
therefore unfeasible.
The design of a feasible version of the GVA must begin, therefore,
by reducing the set of possible types to some set of singly-exponential
size. 
All implementations of auctions assume that the bidders express
their preferences by a small set of {\em bids}.
We shall start with a most sweeping restriction:
in Section~\ref{sec:complexbidders} we shall consider relaxing 
this restriction.

We shall assume that bidders are single-minded and care only about
one specific (bidder-dependent) set of goods: if they do not 
get this set they value the outcome at the lowest possible value: $0$.
In other words, our bidders are restricted to one single bid.
\begin{definition}
\label{def:singleminded}
Bidder $i$ is {\em single-minded} if and only if there is a set \mbox{$s \subset G$}
of goods and a value \mbox{$v \in  {\bf R}_{+}$} such that its type $t$
can be described as:
\mbox{$t(s') = v$} if \mbox{$s \subseteq s'$} and 
\mbox{$t(s') = 0$} otherwise. 
\end{definition} 
We shall denote by \mbox{$\langle s , v \rangle$} the type just described.
Note that a single-minded bidder enjoys {\em free disposal}.
We shall assume, in most of this paper, that all bidders are single-minded,
i.e.\ there are sets of goods $s_{i}$ and non-negative real numbers $v_{i}$
such that bidder $i$ is of type \mbox{$\langle s_{i} , v_{i} \rangle$}.
We shall denote by $\Sigma$ the set of all single-minded types.
The size of the set $\Sigma$, contrary to the size of $\Theta$,
is singly-exponential in $k$. A string of polynomial size will be enough
to code the declarations of the bidders: it will describe a set of goods
and a value. In this setting, we identify bids and bidders.

Note that in a simple auction, i.e.\ \mbox{$k = 1$}, 
a bidder is single-minded if he values at $0$ all allocations in which he
does not obtain the good and at some non-negative value the allocation
in which he gets the good.
So, essentially, in a simple auction, a single-minded bidder is a bidder
who does not care who gets the object if he does not get it, i.e.\
has no externalities.

We shall design a feasible truthful mechanism for combinatorial 
auctions among single-minded bidders.
It may at first seem that this
is a futile exercise, but at least anecdotal evidence suggests
that this single-mindedness is not an uncommon situation. Indeed,
R.~Wilson~\cite{Wilson:private} reports that, in the GVA used for 
selling timber harvesting rights in New Zealand, the bidders were
almost single-minded: they were typically interested in all of the 
locations in a specific geographical area. It might also seem that
this restriction does away with the computational issue; however, as we see 
in the next section, GVAs are unfeasible even with the restriction to
single-minded bidders. 
In Section~\ref{sec:complexbidders}, we shall discuss the generalization
of our results to larger families of bidders.
\section{Unfeasibility of the GVA}
\label{sec:infeas}

Let us now assume that all bidders are single-minded, i.e.\ the set
of all possible types is now $\Sigma$.
It follows easily from Proposition~\ref{prop:GVA} that,
in a GVA, a single-minded bidder of type
\mbox{$\langle s , v \rangle$} never pays more than $v$ and 
pays nothing if he is not allocated the whole set $s$.

In a GVA, the allocation is the one defined in Equation~\ref{eq:f}.
Computing this allocation requires optimizing 
\mbox{$\sum_{i = 1}^{n} d_{i}(a)$} over all $a$'s in the set ${\cal O}$
that is of exponential size. 
One may suspect that this an unfeasible task.
Indeed, the problem of finding the allocation of Equation~\ref{eq:f}
has been shown in~\cite{RothPekHar} to be unfeasible.
We remark that the restriction to single-minded bidders 
does nothing to alleviate the problem.
Some care is needed in describing the NP-hardness result because
we have two parameters to deal with: $k$, the number of goods and $n$,
the number of bidders.
\begin{theorem}
\label{the:NPhard1}
Let a single-minded type 
\mbox{$d_{i} = \langle s_{i} , v_{i} \rangle$}, \mbox{$s_{i} \subseteq G$},
\mbox{$v_{i} \in {\bf R}_{+}$} be given for each bidder \mbox{$i \in P$}.
Let \mbox{$\mid G \mid = k$} and \mbox{$\mid P \mid = n$}.
If $k$ and $n$ grow in a  polynomially related way, 
the problem of finding an allocation $a$ that maximizes 
\mbox{$\sum_{i = 1}^{n} d_{i}(a)$} is NP-hard.
Moreover, the existence of a polynomial time algorithm 
guaranteed to find an allocation
whose value is at least $k^{- 1 / 2 + \epsilon}$ 
times the value of the optimal
solution would imply that NP = ZPP.
\end{theorem}
A short note on complexity classes:
in the above, NP is the class of sets for which membership can be decided 
non-deterministically in polynomial time and ZPP is the sub-class of NP consisting of those sets
for which there is some constant $c$ and a probabilistic
Turing machine $M$ that on input $x$ runs in expected time
\mbox{$O({\mid x \mid}^{c})$} and outputs $1$ if and only if
\mbox{$x \in L$}. The question of whether NP = ZPP is a deep
open question in theoretical computer science, related with the famous
P = NP question. NP = ZPP is not known to imply P = NP, but does imply
NP = RP = co-RP = co-NP. P = NP obviously implies NP = ZPP.
RP is the class of sets for which 
membership can be decided in polynomial time by a randomizing algorithm.
The class co-RP is the class of sets whose complements are in RP: 
non-membership can be decided polynomially by a randomizing algorithm.
Similarly for co-NP. ZPP is the intersection of RP and co-RP.
(end of short note).
\proof 
The problem at hand may be described as the weighted version of the
Set Packing problem of~\cite{Karp:Reducibility}.
Karp shows that Set Packing is NP-hard by reducing the Clique problem to it.
The $k$ used in this reduction is of the order
of $n^2$. 
A direct reduction of Clique to our allocation problem is obtained in the following
way. Given a graph $G$, let the goods be the edges and the bids be the vertices.
Each vertex requests the edges it is adjacent to for a price of $1$.
An optimal allocation is a maximal independent vertex set.
H{\aa}stad~\cite{Hastad:CliqueActa} has shown that Clique
cannot be approximated within \mbox{${\mid V \mid}^{1 - \epsilon}$}
unless NP = ZPP. The reduction mentioned above shows our claim. 
\QED
In Theorem~\ref{the:NPhard1}, we required $k$ and $n$ to grow in a polynomially
related way. This restriction is needed.
On one hand, if \mbox{$n \leq \log k$}, an optimal allocation
may be found in time linear in $k$.
On the other hand, if \mbox{$k \leq \log n$}, 
then dynamic programming provides
an optimal allocation in time quadratic in $n$, as shown in~\cite{RothPekHar}.
 

Let us now consider the significance of Theorem~\ref{the:NPhard1}.
Even if (single-minded) bidders declare their type truthfully,
we cannot always attain an efficient allocation.
Global restrictions on the structure of the set of bidders 
are considered in~\cite{RothPekHar} and shown to allow a polynomial
search for the efficient allocation.
They restrict the possible types of the bidders to a small subset of $\Sigma$,
based on some inherent structure of $G$.
It does not seem those restrictions can be met in practice. 

If the number of goods is large, we may either find an algorithm
that computes the efficient allocation but may, in the worst cases,
never terminate (for all practical purposes) or settle for an algorithm
that provides a sub-efficient allocation.
Both ideas have been proposed in~\cite{FLBS:99,Sandholm99}.
But the impact of such an approximation on the quality of the mechanism,
i.e.\ its truthfulness, or the revenue it generates, has not been studied.
A pioneering study of the properties of approximate mechanisms, but not
for combinatorial auctions, may be found in~\cite{NisanRonen:AMD}.
In section~\ref{sec:greedy}, we shall provide a feasible approximation
algorithm that appears to be very effective in practice and, 
in section~\ref{sec:positive}, we shall describe a payment scheme,
different from the GVA's, that guarantees truthfulness.
The payment scheme is carefully tailored to the specific approximation
algorithm.
 
\section{The greedy allocation}
\label{sec:greedy}
Since an efficient solution seems out of reach, we shall look for
an approximately efficient solution.
We shall propose a family of algorithms that provide such
an approximation.
Each of those algorithms runs in almost linear time in $n$, the number
of single-minded bidders.
One algorithm of the family guarantees an approximation ratio of
$k^{- 1 / 2}$.

A single-minded bidder declaring \mbox{$\langle s , a \rangle$}, with
\mbox{$s \subseteq G$} and \mbox{$a \in {\bf R}_{+}$}
will be said to put out a bid \mbox{$b = \langle s , a \rangle$}.
We shall use $s(b)$ and $a(b)$ to denote the components of $b$
and call $a(b)$ the {\em amount} of the bid $b$.
As explained in section~\ref{sec:single}, we identify bids and bidders.
Two bids \mbox{$b = \langle s , a \rangle$} and
\mbox{$b' = \langle s' , a' \rangle$} conflict if
\mbox{$s \cap s' \neq \emptyset$}.

The algorithms we consider execute in two phases.
\begin{itemize}
\item in the first phase, the bids are sorted
by some criterion. The algorithms of the family are distinguished
by the different criteria they use. Since there are $n$ bids,
this phase takes time of the order of \mbox{$n \log n$}.
We assume a criterion, i.e.\ a norm is defined and 
the bids are sorted in decreasing
order following this norm.
Since we shall have, in Theorem~\ref{the:main}, to compare the sorted lists of
bids of slightly different auctions, we also assume a consistent treatment
of ties, i.e., bids with equal norms.
Formally, we shall assume that no two different bids have
the same norm, i.e., there are no ties.
\item in the second phase, a greedy algorithm generates an allocation.
Let $L$ be the list of sorted bids obtained in the first phase.
The first bid of $L$, say \mbox{$b = \langle s , a \rangle$}
is granted, 
i.e.\ the set $s$ will be allocated to $b$ and then the algorithm examines
each bid of $L$, in order, and grants it if it does not conflict
with any of the bids previously granted.
If it does, it denies, i.e.\ does not grant, the bid.
This phase requires time linear in $n$.
\end{itemize}
The use of such a greedy scheme is very straightforward and speedy.
We shall now discuss its efficiency: how efficient is the allocation
generated?
The efficiency of the allocation generated depends obviously
both on the criterion used in the first phase and on the types of the
bidders,
or on the distribution with which the bidders are generated.
It is clear that, to obtain allocations close to efficiency, one should
use a norm that pushes bids that have a good chance to be part
of an efficient allocation toward the beginning of the list $L$.
The amount of a bid is a good criterion in this respect: we want
bids with higher amounts to have a larger norm than bids with
lower amounts,
at least when the bids are for the same set of goods.
Similarly, leaving the amount of a bid unchanged but making its bundle
a smaller set (inclusion-wise), should also increase the norm.
We shall require that changing $s$ to $s'$ with \mbox{$s' \subset s$}
or changing $v$ to $v'$ with \mbox{$v' > v$} increase the norm
of a bid. Let us call this property {\em bid-monotonicity}.
This is the only requirement we shall make.
Many criteria satisfy it.

In real-life situations, one can typically find a suitable natural norm
related to the economic parameters of the bundle that measures the
a-priori attractiveness of the bid (for the auctioneer).
In the FCC auction, goods (licenses) are characterized by the
population they cover. The (inverse of the) sum of those populations is a 
good indicator.
In the abstract, if we know nothing concrete about the goods,
our best bet is to use the size of the set of goods mentioned in a bid.
We shall look in particular at the average-amount-per-good measure.
\begin{definition}
\label{def:avamount}
The average amount per good of a bid 
\mbox{$b = \langle s , a \rangle$} is \mbox{${a} \over {\mid s \mid}$}.
\end{definition}
Sorting the list $L$ by descending average amount per good is a very
reasonable idea.
But many other possibilities may be considered.
Sorting $L$ by descending amounts for example, or, more generally
sorting $L$ by a criterion of the form
\mbox{${a} \over {{\mid s \mid}^l}$} for some number $l$, \mbox{$l \geq 0$},
possibly depending on $k$.
All such criteria satisfy bid-monotonicity.

How good is the greedy allocation in comparison with the optimal
one? For $l = 1$, the worst case may be analyzed without much difficulty.
The ratio between the total value of the optimal allocation
and that of  the allocation found by the greedy algorithm cannot be larger
than $k$, and this bound is tight.
As usual in this sort of situations, on the average, on realistic
distributions of bids, the performance of the greedy allocation scheme
is much better than the lower bound above.
We have been able to perform a full analysis of the worst case performance
of those norms for different $l$'s and found out that \mbox{$l = 1 / 2$}
is best: it guarantees an approximation ratio of at least 
\mbox{$\sqrt{k}$} and, by Theorem~\ref{the:NPhard1}, this is, up to
a multiplicative constant, essentially, the best approximation ratio 
one can hope for a polynomial-time algorithm.
The \mbox{$\sqrt{k}$} upper-bound improves on the previously best known
result 
of~\cite{Halldors:COCOON99} by a factor of $2$.
The following has, since, been generalized to multi-unit combinatorial
auctions in~\cite{GonLeh:EC00}.
\begin{theorem}
\label{prop:opt}
The greedy allocation scheme with norm 
\mbox{${a} \over {\mid s \mid}^{1 / 2}$} approximates the optimal
allocation within a factor of \mbox{$\sqrt{k}$}.
\end{theorem}
\proof
Assume the bids (i.e., bidders) are \mbox{$\langle s_{i} , a_{i} \rangle$}
for \mbox{$i = 1 , \ldots , n$}. Let \mbox{$w_{i} = \mid s_{i} \mid$}.
Our norm is: \mbox{$r_{i} = a_{i} / \sqrt{w_{i}}$}.
Let $OP$ be the optimal solution, i.e., the set of bids contained
in the optimal solution.
The value of the optimal solution is \mbox{$\alpha = \sum_{i \in OP} a_{i}$}.
Let $GR$ be the solution obtained by the greedy allocation and $\beta$
its value: \mbox{$\beta = \sum_{i \in GR} a_{i}$}.
We want to show that:
\begin{equation}
\label{eq:goal}
\alpha \leq \beta \: \sqrt{k}.
\end{equation}
Notice, first, that we may, without loss of generality, assume that
the sets $OP$ and $GR$ have no bid in common.
Indeed, if they have, one considers the problem in which the common bids and
all the units they request have been removed. The greedy and optimal solutions
of the new problem are similar to the old ones and the inequality for the new smaller
problem implies the same for the original problem.

Let us consider $\beta$.
By elementary algebraic considerations:
\[
\beta = \sum_{i \in GR} a_{i} \geq \sqrt{\sum_{i \in GR} {a_{i}}^{2}} =
\sqrt{\sum_{i \in GR} {r_{i}}^{2} \: w_{i}} 
\]
Consider $\alpha$.
By the Cauchy-Schwarz inequality:
\[
\alpha = \sum_{i \in OP} r_{i} \: \sqrt{w_{i}} \leq \sqrt{\sum_{i \in OP} {r_{i}}^{2}}
\: \sqrt{\sum_{i \in OP} w_{i}}.
\]
The expression \mbox{$\sum_{i \in OP} w_{i}$} represents the total number
of goods allocated in the optimal allocation $OP$
and is therefore bounded from above by $k$, the number of goods available.
We conclude that:
\[ 
\alpha \leq \sqrt{\sum_{i \in OP} {r_{i}}^{2}} \: \sqrt{k}.
\]

To prove~(\ref{eq:goal}), it will be enough, then, to prove that:
\[
\sum_{i \in OP} {r_{i}}^{2} \leq 
\sum_{i \in GR} {r_{i}}^{2} \:  w_{i}.
\]

Consider the optimal solution $OP$. By assumption, the bids of $OP$ did not
enter the greedy solution $GR$. This means that, at the time any such bid $i$
is considered during the execution of the greedy algorithm, it cannot
be entered in the partial allocation already built.
This implies that there is a good \mbox{$l \in s_{i}$} that has already been 
allocated in the partial greedy solution, i.e., there is a bid $j$ in $GR$,
with \mbox{$r_{j} \geq r_{i}$} and \mbox{$l \in s_{j}$}. 

A number of different bids from $OP$ may, in this way be associated with the
same bid $j$ of $GR$, but at most $w_{j}$ different bids of $OP$ may be
associated with bid $j$ of $GR$, since the sets of goods requested by two different 
bids of $OP$ have an empty intersection.
If $OP_{j}$ is the set of bids of $OP$ that are associated with bid $j$:
\[
\sum_{i \in OP_{j}} {r_{i}}^{2} \leq {r_{j}}^{2} \:  w_{j}.
\]
\QED
In other words, the greedy scheme does not guarantee any fixed ratio
of approximation, but guarantees the best achievable ratio 
(assuming NP $\neq$ ZPP).
Experiments reported about in~\cite{GonLeh:EC00} have confirmed that,
on average for a specific distribution, the greedy algorithm using the norm
of Theorem~\ref{prop:opt} performs extremely well, much better than the
lower bound described in the Theorem.
More experiments are necessary to study the average
case performance of different norms. 
In the sequel, all examples will use the average amount per good
criterion but it is not difficult to find similar examples for
any criterion of the form
\mbox{${a} \over {{\mid s \mid}^l}$} .

\begin{example}
\label{ex:greedyall}
Assume there are two goods $a$ and $b$ and three bidders 
Red, Green and Blue.
Red bids $10$ for $a$, Green bids $19$ for the set $\{a , b \}$ 
and Blue bids $8$ for $b$.
We sort the bids by decreasing average amount and obtain:
Red's bid for $a$ (average $10$), 
Green's bid for  $\{a , b \}$ (average $9.5$) and 
Blue's bid for $b$ (average $8$).
The greedy algorithm grants Red's bid for $a$, denies Green's bid for
$\{a , b \}$ since it conflicts with Red's and grants Blue's bid
for $b$. The allocation is not efficient. The efficient allocation
grants Green's bid for $\{a , b \}$ and denies both other bids.
\end{example}


Our goal is to devise truthful mechanisms for combinatorial
auctions among single-minded bidders. 
Given a suitable greedy allocation, can one find a payment scheme
that makes the pair a truthful mechanism?

\section{Greedy allocation and Clarke's payment scheme do not make a truthful
mechanism, even for single-minded bidders}
\label{sec:neg2}
In section~\ref{sec:positive}, a mechanism based
on the greedy allocation will be built and shown to be truthful if
all bidders are single-minded.
In this section, we show that the use of Clarke's payment scheme,
used in the GVA
and described in Equation~\ref{eq:p},
in conjunction with the greedy allocation does {\em not} make for a truthful
mechanism, even if bidders are single-minded.
In other terms, if the greedy allocation 
and Clarke's payment scheme are used,
a bidder may have an incentive to lie about his valuation.
The payment scheme used in the truthful
mechanism of section~\ref{sec:positive} is different from Clarke's.
This is in stark contrast with the almost universal use of Clarke's scheme
for devising mechanisms that are truthful in dominant strategies.
Even in~\cite{NisanRonen:AMD} where approximate mechanisms are shown
to be truthful, the payment schemes are Clarke's scheme.

A very simple example will suffice. 
\begin{example}
\label{ex:Clarke}
As in Example~\ref{ex:greedyall},
there are two goods $a$ and $b$ and three bidders Red, Green, and Blue.
Red bids $10$ for $a$, Green bids $19$ for the set $\{a , b \}$ and
Blue bids $8$ for $b$. 
The greedy algorithm grants Red's and Blue's bids and denies
Green's bid, i.e.\ \mbox{$f(D)(a) = Red$} and \mbox{$f(D)(b) = Blue$}. 
We shall now compute Red's payment.
For this allocation we have the following declared valuations:
\mbox{$v_{Blue} = 8$} and \mbox{$v_{Green} = 0$}.
If Red had bid zero, the greedy algorithm would have granted Green's bid and
denied Blue's bid.
Therefore,  the allocation $f(Z)$ is defined by:
\mbox{$f(Z)(a) = f(Z)(b) = Green$}, where
\mbox{$v_{Blue} = 0$} and
\mbox{$v_{Green} = 19$}.
Clarke's payment scheme gives to Red: $8-0$ for Blue and $0-19$ for Green,
i.e.\ Red pays $11$.
Red ends up paying more than the amount he declared.
If Red has been truthful and his valuation is indeed $10$, his utility is 
$-1$. He would have been better off lying, under-bidding at, 
say $9$, or $0$.
In such a case the greedy algorithm would have granted Green's bid and denied
Blue's and Red's bids and the payment to Red would have been zero,
making his utility $0$, better than $-1$.
\end{example}
Since this example is very simple and can be embedded in many more
complex situations, we may conclude that, typically, the use of
a method that is only approximately efficient is incompatible with the
use of a Clarke's payment scheme.
The next sections present a positive result: there is a payment scheme
(necessarily different from Clarke's) that makes truth-telling a
dominant strategy.

\section{A sufficient condition for a truthful mechanism for single-minded
bidders}
\label{sec:axiomatics}


We shall describe in this section a number of properties of allocation
schemes and of payment schemes for combinatorial auctions. 
Those properties seem natural properties
to expect from a truthful mechanism and they are
satisfied by the GVA. 
We shall then show that any mechanism that satisfies those properties
is truthful.
In the literature, incentive-compatibility (i.e., truthfulness) seems
to have been considered only in connection with efficient mechanisms,
i.e., mechanisms that allocate the goods in an optimal way 
(see~\cite{KrishnaPerry:eff,MondTenn:asympt} for example).
The conditions presented here are remarkable 
in that they apply to non-efficient
mechanisms too.
In section~\ref{sec:positive}, we shall describe a payment scheme
and show that the greedy allocation scheme, together with this new
payment scheme, satisfy those properties.
The properties we are about to describe concern combinatorial auctions
among single-minded bidders. The question of generalizing those
conditions to a more general setting is an intriguing one.
Independently of this work, such a setting has been proposed
in~\cite{KfirMondTen:private}. Their setting is rich enough to
encompass combinatorial auctions among single-minded bidders, but
not among arbitrary bidders. Our mechanism does not satisfy their 
Axiom 2 and its payment scheme is not of the Clarke's type they propose.
The properties below are sufficient conditions for truthfulness
and we do not claim they are necessary. Some of them are obviously
not necessary. Nevertheless many of those properties can be shown to
be necessary in the presence of others and for some others one can show that
given any truthful mechanism one can easily describe another similar truthful
mechanism that satisfies them. We leave to further work the exact 
characterization of truthful mechanisms for combinatorial auctions among
single-minded bidders.

The general structure of the properties of interest is that they
consider a given set of single-minded types and vary {\em one} of those types.
They restrict the changes that can appear in the allocation or the
payments as a result of such a change.
Let declarations be fixed, but arbitrary, for all bidders except $j$. 
Consider two possible declarations for $j$: 
\mbox{$\langle s , v \rangle$} and \mbox{$\langle s' , v' \rangle$}. 
Given an allocation scheme $f$ and a payment scheme $p$,
we shall consider the allocations and payments generated by both
declarations of $j$. 
Let $g_{i}$ be the set of goods obtained by bidder $i$ 
if $j$ declares \mbox{$\langle s , v \rangle$}, and
$g'_{i}$ the set he obtains if $j$ declares \mbox{$\langle s' , v' \rangle$}.
Similarly denote by $p_{i}$ and $p'_{i}$ the payments of $i$.

Our first property requires that the allocation, among single-minded
bidders, be exact, i.e.\ a single-minded bidder either gets exactly
the set of goods he desires, nothing added, or he gets nothing.
He never gets only part of what he requested.
This is a very natural property, when dealing with single-minded bidders:
the valuation of the bidder does not increase by giving him part of 
what he requested instead of nothing 
or by giving him more than what he requested instead of just the bundle
he requested.
\[
{\bf Exactness} \ \ \ \ \ \ \ \ 
{\rm Either} \ g_{j} = s \ {\rm or} \ 
g_{j} = \emptyset
\]
In an exact allocation, we shall say that $j$'s bid is {\em granted} in
the first case, and {\em denied} in the second case.
In such a scheme, the allocation may be viewed as a
set of bids (or bidders) that is conflict-free, i.e.\ the $s$ coordinates
have pairwise empty intersections.
A GVA, as we defined it, does not in fact always satisfy Exactness.
If nobody is interested in $a$, an optimal allocation could still
allocate it to one of the bidders. An obvious modification of the GVA
{\em for single-minded bidders} can ensure Exactness.

Our next property, Monotonicity, 
also concerns only the allocation scheme.
It requires that, if $j$'s bid is granted if he declares 
\mbox{$\langle s , v \rangle$}, it is also
granted if he declares \mbox{$\langle s' , v' \rangle$} 
for any \mbox{$s' \subseteq s$},
\mbox{$v' \geq v$}.
In other words: proposing more money for fewer goods 
cannot cause a bidder to lose his bid.
It follows that, similarly, offering less money for more goods 
cannot cause a lost bid to win.
Formally:
\[
{\bf Monotonicity} \ \ 
s \subseteq g_{j} , \
s' \subseteq s , \ v' \geq v \ \Rightarrow \ s' \subseteq g'_{j}
\]
The GVA's allocation scheme picks the efficient allocation, i.e.\
the allocation that maximizes the sum of the amounts of 
a conflict-free subset of bids. If a bid is included in the optimal
allocation and its amount increases then the same allocation's total
amount increases by the same amount and therefore stays optimal.
Similarly, if the amount stays unchanged but the set of goods requested
becomes smaller (inclusion-wise), the previous allocation,
after the obvious change, is still conflict-free and its total
amount has not changed. Any allocation not containing the new bid 
was a suitable allocation before the change and therefore is not better.
Similarly, if a bid is denied and its amount decreases, the optimal
allocation's value stays fixed but the value of any allocation including
the bid decreases, and similarly when varying the set $s$.
We conclude that, assuming that there is a unique optimal allocation, 
the GVA's allocation scheme satisfies Monotonicity.
In general, when many allocations could be tied for optimality,
a GVA scheme may not be monotonic, but one may may modify the GVA
scheme to ensure Monotonicity.

We must immediately consider the consequences of Monotonicity,
since we shall need them in stating the upcoming Critical property.
\begin{lemma}
\label{le:critical}
In a mechanism that satisfies Exactness and Monotonicity,
given a bidder $j$, a set $s$ of goods and declarations 
for all other bidders,
there exists a critical value $v_{c}$ such that
\begin{eqnarray*}
\forall v , v < v_{c} \ \Rightarrow \ g_{j}  = \emptyset ,
\\
\forall v , v > v_{c} \ \Rightarrow \ g_{j}  = s ,
\end{eqnarray*}
\end{lemma}
We allow $v_{c}$ to be infinite if 
\mbox{$f(A^{s , v})^{-1}(j) = \emptyset$} for every $v$.
Note that we do not know whether $j$'s bid is granted or not in case
\mbox{$v = v_{c}$}.
\proof
By Monotonicity, the set of $v$'s for which \mbox{$g_{j} = \emptyset$} is
empty (in which case take \mbox{$v_{c} = 0$}), 
of the form \mbox{$[0 , v_{c}[$}, 
of the form \mbox{$[0 , v_{c}]$} or equal to ${\cal R}_{+}$.
\QED

Our third property deals with a satisfied bidder:
a satisfied bidder pays exactly the critical value of Lemma~\ref{le:critical},
i.e.\ the lowest value he could have declared and still be allocated the
goods he desires.
\[
{\bf Critical} \ \ \ \ \ \ \ s \subseteq g_{j} \ \Rightarrow \ 
p_{j} = v_{c}
\]
Notice that Critical says, first, that the payment for a bid that is
granted does not depend on the amount of the bid, it depends only on the other
bids. Then it says that it is exactly equal to the critical value
below which the bid would have lost.

Critical is a necessary property for a truthful mechanism that satisfies
Exactness, Monotonicity and the Participation property below.
If the payment $p$ is smaller than $v_{c}$, any bidder with real value between
$p$ and $v_{c}$ looses if he declares truthfully but wins and pays less than 
his true value if he declares just above $v_{c}$.
If the payment $p$ is larger than $v_{c}$, any bidder with real value
between $v_{c}$ and $p$ wins but gets negative utility if he declares
truthfully and would be better off declaring a value below $v_{c}$ and loosing.
Since a GVA is truthful and satisfies Exactness, Monotonicity and 
Participation, it also satisfies Critical.


 
Our last property concerns the payment scheme.
Together with Critical, it implies that the utility of no truthful 
bidder is negative.
It concerns unsatisfied bidders, i.e.\ bids that are denied.
We require that an unsatisfied bidder pay zero.
The utility of an unsatisfied bidder is then zero.
This is simply tuning the utility scales of the different bidders,
or, ensuring that bidders may not loose by participating in the auction.
\[
{\bf Participation} \ \ \ \ \ \ \ s \not \subseteq g_{j} \ \Rightarrow \ 
p_{j} = 0
\] 
A GVA satisfies Participation. In fact the second term of Equation~\ref{eq:p}
is precisely tuned to satisfy Participation.

Any mechanism that satisfies the conditions 
above is truthful.
A number of preliminary lemmas are needed.
\begin{lemma}
\label{le:denied}
In a mechanism that satisfies Exactness and Participation, a bidder
whose bid is denied has utility zero.
\end{lemma}
\proof
By Exactness, the bidder gets nothing and his valuation is zero.
By Participation his payment is zero.
\QED
\begin{lemma}
\label{le:truthleq0}
In a mechanism that satisfies Exactness, Monotonicity, Participation 
and Critical a truthful
bidder's utility is non-negative.
\end{lemma}
\proof
If $j$'s bid is denied, we conclude by Lemma~\ref{le:denied}.
Assume $j$'s bid is granted and his type is 
\mbox{$\langle s , v \rangle$}.
Since he is truthful, his declaration is 
\mbox{$d_{j} = \langle s , v \rangle$}.
We conclude that $j$ is allocated $s$ and his valuation is $v$.
By Lemma~\ref{le:critical}, since $j$'s bid is granted,
\mbox{$v \geq v_{c}$}.
By Critical, $j$'s payment is $v_{c}$, and his utility is
\mbox{$v - v_{c} \geq 0$}.
\QED
The next lemma shows that a bidder cannot benefit from lying
just about his value (he truthfully declares the set of goods he is interested
in).
\begin{lemma}
\label{le:truthvalue}
In a mechanism that satisfies Exactness, Monotonicity, Participation and 
Critical,
a bidder $j$ of type \mbox{$\langle s , v \rangle$}
is never better off declaring \mbox{$\langle s , v' \rangle$}
for some $v' \neq v$ than by being truthful.
\end{lemma}
\proof
Compare the case $j$ bids, truthfully, \mbox{$\langle s , v \rangle$}
and the case he bids \mbox{$\langle s , v' \rangle$}.
Let $g_{j}$ be the bundle he gets in the first case and $g'_{j}$ the bundle
he gets in the second case.
If $j$'s bid is denied in the second case, i.e.\ if
\mbox{$g'_{j} \neq s$}, then,
by Lemma~\ref{le:denied} his utility is zero in the second case 
and by Lemma~\ref{le:truthleq0} his utility in the first case is non-negative.
The claim holds.

Assume therefore that \mbox{$g'_{j} = s$}.
If both bids are granted, $j$ has the same valuation ($v$) and pays the
same payment, $v_{c}$ (by Critical).
If \mbox{$g'_{j} = s$} but \mbox{$g_{j} = \emptyset$},
it must be the case that \mbox{$v \leq v_{c} \leq v'$}.
Being truthful gives $j$, by Lemma~\ref{le:denied}, zero utility.
Lying gives him utility \mbox{$v - v_{c} \leq 0$}.
\QED
\begin{lemma}
\label{le:P-mon}
In a mechanism that satisfies Exactness, Monotonicity and Critical,
a bidder $j$ declaring type \mbox{$\langle s , v \rangle$} 
whose bid is granted,
i.e.\ \mbox{$g_{j} = s$}, pays a price $p_{j}$ that is at least
the price $p'_{j}$ that he would have paid had he declared his type as
\mbox{$\langle s' , v \rangle$} for any \mbox{$s' \subseteq s$}.
\end{lemma}
\proof
By Monotonicity, the bid \mbox{$\langle s' , v \rangle$}
would have been granted and by
Critical, the price $p'_{j}$ paid for such a bid 
satisfies: for any \mbox{$x < p'_{j}$}
the bid \mbox{$\langle s' , x \rangle$} would not have been granted.
By Monotonicity, for any such $x$ the bid \mbox{$\langle s , x \rangle$}
would not have been granted.
By Critical, for any $x$ such that \mbox{$x > p_{j}$}, the bid
\mbox{$\langle s , x \rangle$} would have been granted.
We conclude that \mbox{$p'_{j} \leq p_{j}$}.
\QED
Finally we may prove a central result.
\begin{theorem}
\label{the:pties}
If a mechanism satisfies Exactness, Monotonicity, 
Participation and Critical,
then it is a truthful mechanism.
\end{theorem}
\proof
Suppose $j$'s type is \mbox{$\langle s , v \rangle$}. 
Could $j$ have any interest in declaring
his type as \mbox{$\langle s' , v' \rangle$}?
By Lemma~\ref{le:truthleq0} the only case we have to consider is when declaring
\mbox{$\langle s' , v' \rangle$} $j$ gets a positive utility, 
and by Lemma~\ref{le:denied}
this means that $j$'s bid is granted.
Assume, therefore that \mbox{$g'_{j} = s'$}.
If \mbox{$s \not \subseteq s'$}, the valuation of $j$ is zero.
Since, by Critical, his payment is non-negative, his utility 
cannot be positive.
Assume then \mbox{$s \subseteq s'$}. Since $j$'s valuation for $s'$ is the
same as for $s$, Lemma~\ref{le:P-mon} implies that, 
instead of declaring \mbox{$\langle s' , v' \rangle$},
$j$ would not have been worse-off by declaring \mbox{$\langle s , v' \rangle$}.
Lemma~\ref{le:truthvalue} implies that declaring 
\mbox{$\langle s , v' \rangle$} cannot be better
than being truthful.
\QED

\section{A truthful mechanism with greedy allocation}
\label{sec:positive}
We shall now describe the payment mechanism that we propose to be used
in conjunction with the greedy allocation of section~\ref{sec:greedy}.
The description of the payments is  tightly linked with that of the
greedy algorithm. The computation of the payment is performed in parallel
with the execution of the greedy algorithm and takes time linear in the
number of bidders for each payment. On the whole computing the allocation and
the payments takes time at most quadratic in the number of bids.

We assume that the criterion used is average amount per good, the adaptation
to most other suitable greedy allocations is obvious.
Informally, a bidder pays, per good, the average price proposed by the
first bid in the list $L$ that is denied because of this bid. 
Consider a bid $j$ in $L$. Let $c(j)$ be the average amount per good of $j$.
We shall denote by $n(j)$ the first bid following
$j$ (bids are sorted in decreasing order, i.e.\ \mbox{$c(j) \geq c(n(j))$}) 
that has been denied but would have been granted were it not for
the presence of $j$.
Assume that such a bid exists. 
Notice that such a bid necessarily conflicts with $j$, and therefore:
\[
n(j) = \min \{ i \mid j < i , s(j) \cap s(i) \neq \emptyset ,
\forall l , l < i , l \neq j , l {\rm \ granted \ }
\Rightarrow s(l) \cap s(i) = \emptyset \}.
\]
\begin{definition}[Greedy Payment Scheme]
\label{def:payment}
Let $L$ be the sorted list obtained in the first phase.
\begin{itemize}
\item $j$ pays zero if his bid is denied or if there is no bid $n(j)$,
\item if there is an $n(j)$ and $j$'s bid \mbox{$\langle s , v \rangle$} 
is granted he pays 
\mbox{$\mid s \mid \times \ c(n(j))$}.
\end{itemize}
\end{definition}
We may now state the main result of this paper.
\begin{theorem}
\label{the:main}
The mechanism composed of the greedy allocation and payment schemes 
is truthful for single-minded bidders.
\end{theorem}
\proof
We shall prove that greedy mechanism satisfies Exactness, Monotonicity,
Participation and Critical and use Theorem~\ref{the:pties}.
The description of the greedy allocation scheme makes it clear that every
bid is either granted or denied. The greedy allocation satisfies Exactness.
For Monotonicity, assume that \mbox{$s \subseteq s'$} and that
\mbox{$v \geq v'$} and let $c$ be the norm of
\mbox{$\langle s , v \rangle$}
and $c'$ the norm of \mbox{$\langle s' , v' \rangle$}.
By our assumption concerning norms we have \mbox{$c \geq c'$}.
If we compare the list $L$ and $L'$ obtained 
respectively, we see that, since there are no ties by assumption,
they differ only in that $j$'s bid may have been moved backwards
by the change from \mbox{$\langle s , v \rangle$} to 
\mbox{$\langle s' , v' \rangle$}.
The greedy allocation algorithm performs, i.e.\ grants or denies bids,
in exactly the same way on $L$ and $L'$ until it gets to $j$'s bid in $L$.
Assume $j$'s bid is denied in $L$: there is some bid that conflict with it
that has been granted already. The same bid also conflicts with $j$'s
bid in $L'$ since \mbox{$s \subseteq s'$} and this bid will also be denied.
Similarly if $j$'s bid in $L'$ is granted, no bid granted before conflicts with
it and therefore no bid granted before $j$'s in $L$ conflicts with it either
and $j$'s bid is also granted in $L$.
We have shown that the greedy allocation satisfies Monotonicity.
It is clear from the first part of Definition~\ref{def:payment} that
it satisfies Participation.
For Critical, 
notice that the second part of 
Definition~\ref{def:payment} defines the payment for a bid granted 
at exactly the minimal declared value that would have allowed it to
be granted, $v_{c}$.
Any declared value above \mbox{$\mid s \mid \times \ c(n(j))$}
leaves $j$ before $n(j)$. If there was a bid $i$, \mbox{$j < i < n(j)$}
that would prevent the granting of $j$ displaced in such a way,
$i$ would have to be granted and conflict with $j$.
It is therefore a bid denied in the original allocation, that would have been
granted were it not for $j$, contradicting the fact that $n(j)$ is
the first such bid.
Any declared value below \mbox{$\mid s \mid \times \ c(n(j))$}
guarantees the denial of $j$ because $n(j)$ is granted.
\QED

Let us now describe this payment scheme on two examples.
\begin{example}
\label{ex:greedyp}
Consider the bidders of Example~\ref{ex:Clarke}.
The goods are $a$ and $b$ and the bidders are Red, Green, and Blue.
Red bids $10$ for $a$, Green bids $19$ for the set $\{a , b \}$ and
Blue bids $8$ for $b$.
We have seen that Red's and Blue's bids are granted,
Green's bid is denied. This is not the efficient solution.
If Red had not participated, Green's bid would have been
the one with highest average price and would have been granted.
Red pays Green's average price.
Red pays $9.5$.
Green pays $0$, since his bid is denied.
Blue pays $0$ since he is not keeping any other bid from being granted.

Note that a GVA would have allocated both goods to Green and made him
pay $18$.
\end{example}
\begin{example}
\label{ex:greedycomp}
Assume, as usual, two goods and three bidders.
Red bids $20$ for $a$, Green bids $15$ for for $b$ and
Blue bids $20$ for the set $\{a , b \}$.
Red's and Green's bids are granted. Blue's bid is denied.
If Red had not participated, Blue's bid would still have been denied,
because of Green's. Therefore Red pays zero.
Similarly Green pays zero.
Notice that, in this case, the allocation is the efficient allocation,
as in a GVA,
but the GVA's payments are different: Red pays $5$ and Green pays $0$.
\end{example}
\section{Complex bidders}
\label{sec:complexbidders}



Our assumption that bidders are single-minded 
seems very restrictive, is there a way to extend
our results to more complex players?
Why not view a player as a collection of
single-minded agents, or, equivalently, view the type of
a player as a collection of bids?
In such a setting, the game played becomes much richer in strategies
and players may be better-off lying on some of their bids to
obtain an advantage on others.
Our discussion will, by necessity, be sketchy.

In section~\ref{sec:single}, we presented single-minded bidders as
an answer to the combinatorial explosion in bidders' types triggered
by a growth in the number of goods, $k$.
The set of types is doubly-exponential in $k$, but the set of
single-minded types is only exponential in $k$.
In trying to overcome the limitation to single-minded bidders
one could consider any super-set of the single-minded types that
grows only exponentially with $k$.
A very natural idea is to consider players that send off single-minded
agent bidders to do their work. The agents play rationally, but
individually, and bring the goods and the payments due to the player. 
In the final analysis, a player gets all the goods
obtained by each of his agents and pays all the payments imposed
on each of his agents.
A player's strategy is then a {\em small}, i.e.\ polynomial in $k$,
set of single-minded agents, i.e.\ bids.

Such an idea fits very well with ideas popular in the Distributed 
Artificial Intelligence community concerning the role of Intelligent Agents.
In the setting of combinatorial auctions, AI authors 
(\cite{Sandholm99,FLBS:99}) like to consider bidders as placing bids.
Each bid is then adjudicated separately.
Our proposal is a formalization of this idea and enables us to raise
fundamental game-theoretic questions about this setting.
This setting is by no means a trivial restriction.
Notice, for example, that, even though a GVA may be described in terms
of bids placed by the players, a player placing one bid 
for each subset of the goods, 
the allocation and payment schemes 
require knowledge of the identity of the player who placed the bid:
a player can have at most one of his bids granted and his payment is not
a function only of the bids but also of their owners.

One may ask the following questions: 
given a type $t$, not necessarily single-minded, what is a truthful
description of $t$ as a small set of single-minded bidders?
For which types is there such a description?
Given a mechanism, what is the declaration, i.e.\ small set of single-minded
bidders, that a player of type $t$ should use to get the most
out of the mechanism?
Is there a mechanism for which a truthful declaration is
a dominant strategy?
The sequel will show that, if the mechanism uses any reasonable
variation on the greedy allocation, the answer is negative for any
reasonable definition of a truthful description.

First, let us remark that one positive result has been obtained.
Theorem~\ref{the:main} shows that a single-minded bidder has,
in our mechanism, a weakly-dominant strategy that is to tell the truth,
{\em even if the other players are complex players represented by
a collection of single-minded agents}.
But what is the optimal strategy of a complex player, i.e.\ which
agents should he send off?

It is not clear what are the mechanisms we should consider in this 
setting. One could assume a blind mechanism, in which the auctioneer
has to allocate the goods between the single-minded agents without
knowing which agents are owned by the same player.
But one could also provide the auctioneer with this information.
This would allow him, for example, to avoid making the payment
for a bid depend on another bid from the same player, which is certainly 
a step toward truthfulness.
One could also require the auctioneer does not grant more 
than one bid from each bidder, but the literature does not seem to favor
this policy.
As noticed in section~\ref{sec:compl}, a player may naturally express 
complementarity by the bids he puts out, but expressing substitutability
is more difficult.
To this effect, one could allow the players to declare not only a set
of bids but also an incompatibility list describing which of his
bids may not be granted simultaneously.
This is the policy proposed in~\cite{FLBS:99} under the name {\em dummy goods}.

A further discussion of these issues can be left for a future paper
since our result, concerning the greedy allocation's properties, 
is negative and based on a simple situation that can
be embedded in any of the proposals above.
In section~\ref{sec:neg1}, a strong result will be presented but it
is necessarily formal, and general.
Here, we shall present a concrete example.
\begin{example}
\label{ex:concrete}
The mechanism we consider is the greedy mechanism.
Red is a single-minded bidder and his type is 
\mbox{$\langle \{a\} , 12 \rangle$},
i.e.\ he bids $12$ for $a$ alone.
Green is a complex bidder. His type $t_{G}$ is described by:
\mbox{$t_{G}(\{a\}) = 10$}, \mbox{$t_{G}(\{b\}) = 10$} and
\mbox{$t_{G}(\{a , b \}) = 30$}.
Notice that Green exhibits complementarity: he values the set 
\mbox{$\{a , b\}$} at more than the sum of his values for $a$ and $b$.
Whatever stance one takes about the way a set of single-minded bidders
can, in general, 
represent a type, in this case, the set of three 
bids: \mbox{$\langle \{a\} , 10 \rangle$}, 
\mbox{$\langle \{b\} , 10 \rangle$} and 
\mbox{$\langle \{a , b\} , 30 \rangle$} is
a truthful representation of Green's type.
Even if the rules of the auction allow the auctioneer to grant
Green both his bid for $a$ and his bid for $b$, Green cannot complain,
in such a case,
about his bid for the set \mbox{$\{a , b\}$} being denied since he
will, under any reasonable payment scheme, pay less for $a$ and $b$
separately than for his bid for the whole set.
Suppose Green bids truthfully.
The greedy mechanism grants Green's bid for the set \mbox{$\{a , b\}$}
and denies all other bids. Green pays $24$ 
(in a GVA he would pay only $12$), and therefore his utility is $6$.
To eliminate all doubts about the legitimacy of the payment scheme here,
notice that Green's payment is determined by Red's bid, not by Green's 
other bids.

But consider what would have happened if Green had under-bid and declared:
\mbox{$\langle \{a\} , 10 \rangle$}, 
\mbox{$\langle \{b\} , 10 \rangle$} and 
\mbox{$\langle \{a , b\} , 23 \rangle$}.
The greedy mechanism now allocates $a$ to Red (he pays $11.5$) and
$b$ to Green. Green pays zero. His utility is $10$.
Green is better off lying.
Notice that, by lying on his valuation for the set \mbox{$\{a , b\}$},
Green loses ($6$) on this bid: considered in isolation, this bid had
no incentive to lie, but this lie favors the bid for $b$ which happens
to be Green's also.
\end{example}

Example~\ref{ex:concrete} above exhibits a situation in which a gang
of single-minded players may be globally better off under-bidding and losing
utility on one of its bids, in order to have another of the gang's bid
granted and making up for the loss, and more.
A similar situation can arise in which a gang may be better off over-bidding
on a bid $b_{1}$ to ensure that it is granted, even at a loss, 
in order to keep another bidder from getting goods included in another bid
of the gang.

The greedy mechanism is not truthful for complex players.
In the next section we shall show that the fault does not lie with the
payment scheme: no payment scheme can make the greedy allocation 
algorithm truthful. The problem lies with the allocation scheme.
Nevertheless, the greedy mechanism has some truthfulness in it.
If a player's bidding is decided in a myopic way by his single-minded agents 
they will bid truthfully. It is only global considerations that
can induce a society of agents to require its agents not be truthful.
We think we have here some kind of myopic, limited or bounded truthfulness
that may be a very useful ingredient in certain types of mechanisms.
Situations in which the players have too little information and 
too few resources to be able to analyze intelligently 
the global strategic situation may induce them to delegate their
strategy to myopic agents.
In such situations one may be content with a mechanism that
exhibit this kind of limited truthfulness.

\section{No payment scheme makes the greedy allocation a truthful mechanism
for complex bidders}
\label{sec:neg1}
In section~\ref{sec:complexbidders}, we showed that the greedy scheme,
i.e.\ greedy allocation + greedy payment, cannot be extended to
a truthful mechanism for complex players.
We shall now show that no payment scheme can complement the
greedy allocation.

If a bidder is not single-minded, but double-minded, i.e.\ interested
in two different sets of two goods,
there may be no payment scheme that, combined with the greedy allocation
algorithm, will make for a truthful mechanism.
We shall consider a very simple situation: two goods, two bidders,
one of them single-minded, the other one double-minded.
The search for a family of bidders that is significantly larger 
than the single-minded ones and a suitable payment scheme is open,
but it starts with a negative result.
Notice the result does not only show that our payment scheme
is unsuitable, it shows that no payment scheme exists (to be used
in conjunction with the greedy allocation scheme).

Assume there are two goods $a$ and $b$ and two bidders Green and Red.
Red is single-minded and his type is 
\mbox{$\langle \{a\} , 10 \rangle$}.
Red truthfully declares his type.
Green is interested in both $b$ and the set $\{a , b \}$.
His valuation is $0$ for any allocation in which he does not get $b$.
It is $v_{b}$ for any allocation in which he gets $b$ but not $a$,
and it is  $v_{ab}$ if he gets both $a$ and $b$, \mbox{$v_{ab} > v_{b}$}.
Green's declaration is $0$ for all allocations that do not give him $b$,
$g_{b}$ for all allocations that give him $b$ but not $a$ and $g_{ab}$
for the allocation in which he gets both $a$ and $b$.
Notice that four parameters describe the auction:
$g_{ab}$, $g_{b}$, $v_{ab}$ and $v_{b}$.
Assume, furthermore, that \mbox{$0 \leq g_{b} < 10$}.
We reason by contradiction and assume there is a payment scheme
that makes truth-telling a dominant strategy for Green.
Let us consider two cases.

First, assume that \mbox{$g_{ab} > 20$}.
In this case, the greedy algorithm will allocate both goods to Green.
The payment mechanism will make Green pay $p_{ab}$.
Notice that this payment $p_{ab}$ cannot depend on:
\begin{itemize}
\item $g_{ab}$ (as long as \mbox{$g_{ab} > 20$}): 
otherwise, Green would have an interest in declaring the
$g_{ab}$ most favorable to him, irrespective of his $v_{ab}$,
\item $g_{b}$: otherwise, similarly, Green would have an interest in
declaring the $g_{b}$ most favorable to him irrespective of his $v_{b}$,
\item $v_{ab}$: since payments cannot depend on private values,
\item $v_{b}$: similarly.
\end{itemize}
Therefore, $p_{ab}$ is simply a number.
The utility of Green, in this first case, is: \mbox{$v_{ab} - p_{ab}$}.

Consider, now, a second case: \mbox{$g_{ab} < 20$}.
In this case, the greedy algorithm allocates $a$ to Red and $b$ to Green.
Let us denote by $p_{b}$ the payment of Green.
For the same reasons as above, $p_{b}$ cannot be a function
of any of the parameters.
The utility of Green, in this second case, is: \mbox{$v_{b} - p_{b}$}.

Assume that, in fact, Green is bidding his true valuation on $b$,
i.e.\ \mbox{$g_{b} = v_{b}$}.
Since truth-telling is a dominant strategy for Green, it must be the case
that,
\begin{itemize}
\item if \mbox{$v_{ab} > 20$}, Green gets from case 1 not less than 
from case 2, i.e.\ \mbox{$v_{ab} - p_{ab} \geq v_{b} - p_{b} = g_{b} - p_{b}$}
\item if \mbox{$v_{ab} < 20$}, Green gets from case 2 not less than 
from case 1, i.e.\ \mbox{$g_{b} - p_{b} = v_{b} - p_{b} \geq v_{ab} - p_{ab}$}.
\end{itemize}
By considering the case $v_{ab}$ is just greater than $20$ and
$g_{b}$ is just less than $10$, the first inequality gives us
\mbox{$20 - p_{ab} \geq 10 - p_{b}$}, i.e.\ \mbox{$p_{ab} - p_{b} \leq 10$}.
By considering the case $v_{ab}$ is just less than $20$ and
$g_{b}$ is $0$, the second inequality gives us
\mbox{$- p_{b} \geq 20 - p_{ab}$}, i.e.\ \mbox{$p_{ab} - p_{b} \geq 20$}.
A contradiction.


Let us try, now, to discuss the reasons for the negative result just presented.
Why is there a scheme for single-minded bidders and no scheme for more
complex bidders?
The impossibility
to devise a truth-conducing payment scheme around the greedy allocation
stems from the richness of the strategic possibilities offered to a 
complex bidder.
Let us explain why the obvious extension of our payment scheme does not
work. 
Bidder $i$, in order to get good $a$ against the competition of another bidder
interested in \mbox{$\{a , b , c\}$}, may have an interest in over-bidding
on $c$ and get it at a loss, 
just to keep his opponent from getting the whole set.
Similarly, $i$ under-bidding on $a$ and loosing it, may give \mbox{$\{a , b$\}}
to another bidder, which in turn may keep a third bidder from getting
\mbox{$\{b , c\}$} and cause $i$ to get much coveted $\{c\}$.

The discussion just above is, in fact, very similar to Vickrey's discussion
in section~5 of~\cite{Vickrey:61} of the reasons why his scheme
for an auction of identical objects is truth-revealing only if one assumes
buyers of a very simple type: interested in at most one item.

\section{Revenue considerations}
\label{sec:evaluation}
We have described a feasible mechanism for combinatorial
auctions that is truthful when bidders are single-minded.
Should a seller use it for selling goods? 
It is very difficult to say anything general
about the revenue generated by this
mechanism.
We shall compare the revenue generated by our mechanism to the revenue
generated by a GVA.
Since a GVA allocates the goods in an efficient way but our mechanism
does not, one can fear that the revenue generated by our mechanism
will be significantly smaller in all those cases in which the allocation
is not efficient.
This does not seem to be the case.
There are cases in which our algorithm generates a higher revenue than
a GVA and there are cases in which a GVA is preferable.
The comparison does not seem to be tightly correlated to the 
relative efficiency of the allocations.
We shall present four simple situations. 
All examples assume 
single-minded bidders Green, Red, Black and sometimes Blue.
The first two examples are typical of purely combinatorial situations.
\begin{example}
\label{ex:better}
Assume there are four goods, $a$, $b$, $c$ and $d$.
Green is interested in \mbox{$\{a , b \}$}, Red in \mbox{$\{c , d \}$}
and Black in \mbox{$\{a , c \}$}.
All bids are for the same amount: $1$.

Let us first consider a GVA. A GVA allocates the efficient way: 
Green gets \mbox{$\{a , b \}$} and Red gets \mbox{$\{c , d \}$}.
Green and Red pay nothing: if they had not participated only one bid
could have been granted. The revenue generated by a GVA is zero.

Because of the tie our greedy scheme may end in one of three
possible situations, up to symmetry between Green and Red.
First scenario: the order is Green, Red, Black. The allocation
is efficient and nobody pays anything, as for a GVA.
Second scenario: the order is Green, Black, Red. The allocation
is efficient, but this times Green pays $1$, Red pays nothing.
Third scenario: the order is Black, Green, Red. The allocation
is {\em not} efficient: Black gets \mbox{$\{a , c \}$} and $b$ and $d$
are unallocated. Black pays $1$.

In this case, our scheme generates, on average, $2/3$, whereas a GVA
generates $0$.
\end{example}
\begin{example}
\label{ex:notgood}
Four goods, $a$, $b$, $c$ and $d$.
Green is interested in \mbox{$\{a , b \}$}, Red in \mbox{$\{c , d \}$},
Black in \mbox{$\{a , c \}$} and Blue in \mbox{$\{b , d \}$}.
All bids are for the same amount: $1$.

A GVA allocates the efficient way: either to Green and Red or to
Black and Blue. 
In any case each successful bidder pays $1$: the revenue is $2$
and the full surplus is extracted.

Because of the tie our greedy scheme may end in one of three
possible situations, up to symmetry.
First scenario: the order is Green, Red, Black, Blue. The allocation
is efficient (to Green and Red) and nobody pays anything.
Second scenario: the order is Green, Black, Red, Blue. The allocation
is efficient (to Green and Red), 
but this times Green pays $1$, Blue pays nothing.
Third scenario: the order is Green, Black, Blue, Red. The allocation
is efficient (to Green and Red).
Green pays $1$ and Red pays nothing.

In this case, our scheme generates, on average, $2/3$, whereas a GVA
generates $2$.
\end{example}
Our next example is typical of strong complementarity.
\begin{example}
\label{ex:best}
Red bids $20$ for the set \mbox{$\{a , b \}$}, Green bids $9$ for $a$
and Black bids $1$ for $b$.
Both our greedy algorithm and a GVA allocate $a$ and $b$ to Red.
With our scheme Red pays $18$, with a GVA he pays $10$.
\end{example}
\begin{example}
\label{ex:worseeff}
Green bids $20$ for $a$, Red bids $37$ for
the set \mbox{$\{a , b \}$} and Black bids $18$ for $b$.
Both our greedy algorithm and the efficient allocation of the GVA
give $a$ to Green and $b$ to Black.
With us, Green pays $18.5$ and Black pays nothing.
With a GVA, Green pays $19$ and Black pays $17$.
Our mechanism generates $18.5$ to the GVA's $36$.
\end{example}
\begin{example}
\label{ex:worsenoteff}
Green bids $10$ for $a$ and Red bids $19$ for the set \mbox{$\{a , b \}$}.
Our greedy scheme allocates $a$ to Green and leaves $b$ unallocated.
The efficient allocation of the GVA gives both $a$ and $b$ to Red.
In our scheme Green pays $9.5$.
In a GVA, Red pays $10$.
\end{example}
More work is needed to assess the revenue generated by the mechanism
proposed.
\section{Conclusion and future work}
\label{sec:conclusion}
To overcome the complexity of computing the efficient allocation
in combinatorial auctions, we propose to use a greedy approximation
together with a payment scheme tailored to fit it.
The combination provides a truthful mechanism.
This mechanism admits dominant strategies and is therefore very sturdy.

A number of additions, modifications or extensions should be
considered.
Reserve prices are a necessary feature of real-life auctions.
Adding reserve prices to our scheme poses no problem: reserve prices
are bids put out by the auctioneer and truthfulness is still
a dominant strategy for the bidders. 
In a combinatorial auction, the reserve prices can, very naturally,
express the complementarity of the seller.
In particular, a seller who does not want to sell too large sets
of goods to the same buyer, to avoid monopolies for example,
will put high reserve prices for large sets of goods.

Before one can apply the ideas presented here to auctions of identical
items, and to such double auctions, 
those ideas need to be adapted to this setting.
This is the topic of further research.

A combinatorial auction that features a number of different types of goods,
a number of items of each type of goods being for sale, represent
the ultimate combinatorial auction. The ideas presented in this paper
may provide a computationally feasible solution for such auctions.

The revenue generated by the mechanism proposed should be studied
in depth.

The approximation scheme presented in this paper: greedy, is quite
rudimentary. Even though it attains the theoretically optimal (worst-case)
ratio, it should, probably, in practice, be either iterated with
different criteria or be included in some more complex scheme with
some sort of backtracking.
 The main avenue for further research is probably the consideration
such better approximation schemes and the design of suitable 
payment schemes. 
The properties described in section~\ref{sec:axiomatics} are a clear
guide on how to do that. Note, in particular, that Critical leaves
no freedom in the design of the payment scheme.

The properties of section~\ref{sec:axiomatics} are sufficient for
truthfulness, among single-minded bidders,
but some of them also seem to be necessary, at least in the
presence of others.
A full characterization of truthful schemes for combinatorial auctions
should be attempted.

\section*{Acknowledgments}
We want to thank Moshe Tennenholtz and Robert Wilson for their
insightful remarks.

\bibliographystyle{plain}

\end{document}